\documentclass[11pt]{iopart}
\usepackage{bm}
\usepackage{hyperref}
\usepackage{amssymb}
\usepackage{bm}
\usepackage{color}
\usepackage{amsopn}
\usepackage{graphicx}
\renewcommand{\etal}{et al.}

\newcommand{\fnl}{f_{\mathrm{NL}}}

\newcommand{\gnl}{g_{\mathrm{NL}}}

\newcommand{\Mp}{M_{\mathrm{P}}}

\newcommand{\be}{\begin{equation}}
\newcommand{\ee}{\end{equation}}
\newcommand{\N}{M}
\newcommand{\iso}{s}

\renewcommand{\d}{\mathrm{d}}
\newcommand{\vect}[1]{\bm{\mathrm{{#1}}}}
\renewcommand{\e}[1]{\mathrm{e}^{{#1}}}

\renewcommand{\geq}{\geqslant}

\renewcommand{\matrix}[1]{\bm{#1}}

\makeatletter
\newcommand\numberwithin[2]{\@addtoreset{#1}{#2}}
\makeatother
\numberwithin{footnote}{section}

\newcommand{\para}[1]{\vspace{3mm}\par\noindent\textbf{{#1}}}

\begin{document}

	\title{Moment transport equations for the primordial curvature perturbation}
	\author{David J. Mulryne$^1$, David Seery$^2$ and Daniel Wesley$^{3}$}
	\address{\vspace{2mm}
	$^1$ Theoretical Physics Group,	Imperial College, London, SG1 2AS,
	UK\\[2mm]
	$^2$ Astronomy Centre, University of Sussex, Falmer, Brighton,
	BN1 9QH, UK\\[2mm]
	$^3$ Center for Particle Cosmology,
	David Rittenhouse Laboratory \\ University of Pennsylvania,
	209 South 33rd Street, Philadelphia, PA 19104 USA
	}
	\eads{\mailto{d.mulryne@imperial.ac.uk}, \mailto{d.seery@sussex.ac.uk},
		\mailto{dwes@sas.upenn.edu}}
%	\pacs{98.80.-k, 98.80.Cq, 11.10.Hi}
	\begin{abstract}
	In a recent publication, we proposed that
	inflationary perturbation theory
	can be reformulated in terms of a probability transport equation,
	whose moments determine the correlation properties of the
	primordial curvature perturbation.
	In this paper we generalize this formulation to an arbitrary
	number of fields.
	We deduce ordinary differential
	equations for the evolution of the moments of $\zeta$ on superhorizon
	scales, which can be used to obtain
	an evolution equation for the dimensionless bispectrum, $\fnl$.
	Our equations are covariant in field space and allow
	identification of the source terms responsible for
	evolution of $\fnl$.
	In a model with $M$ scalar fields,
	the number of numerical integrations
	required to obtain solutions of these equations
	scales like $\Or(M^3)$.
	The performance of the moment transport algorithm means that
	numerical calculations with $M \gg 1$ fields are straightforward.
	We illustrate this performance with a numerical calculation
	of $\fnl$ in
	Nflation models containing $M \sim 10^2$ fields, finding agreement with
	existing analytic calculations.
	We comment briefly
	on extensions of the method beyond the slow-roll
	approximation, or to calculate higher 
	order parameters such as $\gnl$.
	
	\end{abstract}
	\maketitle
	
	\section{Introduction}
	\label{sec:intro}
	
	In a previous article \cite{Mulryne:2009kh} we introduced a new
	technique (``moment transport'')
	designed to extract the statistics of primordial fluctuations
	generated by inflation.
	Our technique exploits the principle that one never
	uses a model of inflation to predict the properties of any particular
	universe. Instead, one determines averages over
	a large ensemble of universes. Observable quantities are
	predicted on the basis that our universe is
	a typical member of the ensemble.
	Therefore there is no need to study the evolution of fluctuations
	in any specific universe; it is sufficient to compute the evolution
	of the probability density function on the ensemble.
	Ref.~\cite{Mulryne:2009kh} discussed the moment-transport method
	for a restricted range of models.
	In this paper we refine and extend our method to accommodate
	models containing many
	self-interacting scalar fields in a variety of gauges.
	We show that the moment-transport technique yields a system of
	coupled ordinary differential equations which can be used for
	rapid computation of observable quantities, such as the dimensionless
	bispectrum $\fnl$.
	
	Our method is based on
	a transport equation.
	This is a partial differential equation which governs the evolution of a
	probability distribution. In Ref.~\cite{Mulryne:2009kh},
	we showed how a transport equation for a bivariate distribution can be used
	to derive ordinary differential equations which describe
	the evolving moments of any nearly Gaussian distribution.
	Our method was based on diagonalization of the covariance
	matrix describing correlations between the two variables, a process
	analogous to Gram--Schmidt orthogonalization.
	In this paper our first objective
	is to introduce a more elegant method based on
	matrix decomposition.
	Such decompositions generalize immediately to an arbitrary
	number of scalar fields.
	Using the resulting equations we are able to evolve the
	joint probability distribution for any number of variables.

	The moment transport equation approach can 
	be thought of as a reformulation of cosmological perturbation theory on 
	superhorizon scales with the additional virtue that, rather than 
	simply evolving a perturbed quantity, such as $\zeta$, this approach 
	evolves the
	\emph{statistical moments} of a perturbed quantity. This is	
	advantageous because these are the observationally relevant 
	objects.

	The equations presented in Ref.~\cite{Mulryne:2009kh} were
	valid only in the spatially flat slicing.
	In this gauge,
	the joint probability distribution for a collection of light
	scalar fields at horizon-crossing is simple to calculate
	using the methods of quantum field theory
	\cite{Seery:2005gb}.
	Ref.~\cite{Mulryne:2009kh} determined the
	statistics of observational quantities---such
	as the curvature perturbation, $\zeta$, in the uniform
	density gauge---%
	by evolving this distribution to a later time and
	making an appropriate gauge transformation.
	In the present paper,
	our second objective is to generalize the moment-transport method
	to arbitrary slicings of spacetime.
	Once liberated from the restriction to spatially flat slicings
	we are free to calculate the horizon-crossing distribution of the
	curvature perturbation itself, and propagate this
	forward in time. Observational quantities may be
	read directly from this distribution at any desired time,
	with no need for an auxiliary gauge transformation.
		
	Whichever gauge we pick, our predictions for observational
	quantities must agree.
	Nevertheless,
	working directly in the uniform density gauge
	has certain advantages.
	First, slices of constant time have an intuitive and
	unambiguous meaning, being associated with hypersurfaces
	of constant Hubble parameter.
	Second, it is possible to write evolution equations for
	observable quantities such as $\fnl$.
	There is no need to keep track of intermediate, unphysical
	quantities with the associated risk that large cancellations
	occur when taking combinations designed to yield physical observables.
	Third, these evolution equations show explicitly
	which source terms are responsible for the growth and decay of
	the covariance and higher moments of $\zeta$.
	The possibility of
	cancellations obscures this interpretation
	for intermediate variables such as the field perturbations.

	In this paper, we discuss two methods which may be used to extract
	the moment hierarchy associated with a general transport equation.
	In \S\ref{sec:transport} we revisit the technique of Gauss--Hermite
	expansions, employed in Ref.~\cite{Mulryne:2009kh}, which explicitly
	invokes a perturbative expansion around a Gaussian distribution.
	Similar expressions were used by Contaldi \& Magueijo
	to synthesize microwave background maps with a nongaussian component
	\cite{Contaldi:2001wr}, and more recently have been applied to the
	distribution of collapsed structures; among others, see Refs.~%
	\cite{LoVerde:2007ri,Desjacques:2010nn,Chongchitnan:2010xz}.
	Improving the discussion given in Ref.~\cite{Mulryne:2009kh},
	we use a technique of matrix decomposition to find equations
	valid for an arbitrary number of fields. This method
	has several virtues. Most important, in the generalized version to be
	discussed in \S\ref{sec:transport} it is explicitly tensorial.
	Therefore,
	combinatorical factors
	associated with multi-field models are built into the
	formalism and do not need to be addressed directly.

	The Gauss--Hermite
	technique is an example of a cumulant expansion.
	Given a ``kernel'' distribution with a finite number of
	nonzero cumulants, up to some order $M$, we use it as
	a template for
	a general distribution with perturbatively small cumulants of
	order $> M$. For the Gauss--Hermite expansion
	the kernel is a Gaussian with $M = 2$.
	It has zero third- and higher-order cumulants.
	In the absence of special reasons to the contrary,
	the Gauss-Hermite method
	fails when a third- or higher cumulant
	of the probability density function of interest grows to the degree
	that it is not perturbatively small.
		
	In \S\ref{sec:cumulant} we give a different perspective on the
	results of \S\ref{sec:transport}, rederiving them using an alternative
	technique which does not rely on an expansion around
	an unperturbed kernel.
	Introducing generating functions for the moments and cumulants of
	the probability density function of interest, we demonstrate
	the surprising and remarkable fact that the
	\emph{same} transport hierarchy applies for an arbitrary kernel function.
	The disadvantage of this method is a complicated treatment
	of the multi-field combinatorics.
	As well as a check on the correctness of the formulas derived
	in \S\ref{sec:cumulant},
	the method of generating functions clarifies
	under which circumstances
	we can expect the transport hierarchy to apply.
	In particular, it allows a more refined discussion of the error
	involved in truncation of the hierarchy.
	In practical calculations the two methods are essentially equivalent
	because the transport hierarchy must be truncated by assuming
	that an infinite number of cumulants are perturbatively small.

	In \S\ref{sec:inf} we apply our general framework to the fluctuations
	generated by a model of inflation.
	This depends on 
	the selection of a smoothing scale
	and a choice of gauge, which we discuss in
	\S\ref{sec:gauge}.
	The smoothing scale is a measure of
	the spatial scale on which the
	probability density function measures correlations.
	In \S\ref{sec:uniform} we specialize to the uniform density
	gauge and obtain the moment transport equations on this slicing.
	
	Throughout, we adopt units in which $c = \hbar = 1$ but explicitly retain
	the Planck mass, $\Mp = (8\pi G)^{-1/2}$,
	because in some gauges we are obliged to mix
	dimensionless quantities---such as the accumulated e-folds, $N$,
	and its perturbation $\zeta$---with dimensionful ones,
	such as the field values $\phi_i$.
	We label the species of scalar fields by indices $\{ i, j, \ldots \}$.

	\section{Method A: Gauss--Hermite expansions}
	\label{sec:transport}

	In this section, we use the Gauss--Hermite method to
	derive the moment hierarchy associated with a transport equation for
	an arbitrary number of variables $x_i$, where $i = 1, \ldots, N$. In 
	due course these variables will be asociated with cosmological 
	quantities such as field values or curvature, but at present we 
	keep the discussion general.	 
	It will sometimes be convenient to adopt a matrix notation,
	in which the $x_i$ are treated as components of a vector $\vect{x}$.
	In what follows we use matrix and component notation interchangeably.
	
	\subsection{The Gauss--Hermite expansion}
	\label{subsec:gauss}
	We allow $\vect{x}$ to have an arbitrary expectation value
	$\vect{X}(t)$.
	Two-point correlations among the $\vect{x}$ are expressed by the
	covariance matrix $\matrix{\Sigma}(t)$,
	which is defined to satisfy
	\begin{equation}
		\langle (x_i - X_i) (x_j - X_j) \rangle = \Sigma_{ij} .
		\label{eq:cov}
	\end{equation}
	In addition there may be nontrivial third moments,
	$\alpha_{ijk}(t)$,
	\begin{equation}
		\langle (x_i - X_i) (x_j - X_j) (x_k - X_k) \rangle
		=
		\alpha_{ijk} .
		\label{eq:third-moment}
	\end{equation}
	The covariance matrix and third-order moments
	are functions of time,
	as is the position of the centroid $\vect{X}$.
	As in Eqs.~\eref{eq:cov}--\eref{eq:third-moment},
	we will frequently suppress explicit time dependence to avoid
	unnecessary clutter.
	
	Eq.~\eref{eq:cov}
	makes $\matrix{\Sigma}$ a real, symmetric positive-definite
	matrix. It therefore admits a decomposition
	of the form $\matrix{\Sigma} = \matrix{A} \matrix{A}^\mathrm{T}$,
	where $\matrix{A}^\mathrm{T}$ denotes the
	matrix transpose of $\matrix{A}$.
	A candidate decomposition can be obtained by setting
	$\matrix{A} = \matrix{Q} \matrix{\lambda}^{1/2}$
	where $\matrix{\lambda} = \matrix{Q}^T \matrix{\Sigma} \matrix{Q}$ is the
	diagonal
	matrix of eigenvalues of $\matrix{\Sigma}$ and $\matrix{Q}$ is
	orthogonal.
	Other representations may exist.
	For our purpose it is sufficient to pick any one of these
	candidates. The matrix $\matrix{A}$ is only a tool with which to
	construct intermediate quantities and does not occur in the final
	transport equations, so this nonuniqueness does not lead to
	ambiguities.
	Given a choice of $\matrix{A}$, we may define
	standardized variables $z_i(\vect{x})$,
	\begin{equation}
		z_i = A^{-1}_{ij} (x_j - X_j) .
		\label{eq:zdef}
	\end{equation}
	The $z_i$ have zero mean and orthonormal covariances.
	In this and subsequent expressions, the summation convention is
	applied to repeated indices.
	
	If the joint probability distribution of $\vect{x}$ is close to Gaussian
	(an assumption which we relax in Section \ref{sec:cumulant}),
	it may be represented by a Gauss--Hermite expansion
	\begin{equation}
		P(\vect{x}) \, \d^N x
		=
		P_g(\vect{z})
		\left[ 1 + \frac{\alpha^z_{ijk}}{6} H_{ijk} (\vect{z}) \right] 
		\, \d^N x ,
		\label{eq:gh}
	\end{equation}
	where $P_g$ is a normalized Gaussian kernel,
	\begin{equation}
		P_g(\vect{z}) = \frac{1}{(\det 2\pi \matrix{\Sigma})^{1/2}}
		\exp \left( - \frac{\vect{z}^2}{2} \right) .
	\end{equation}
	Note that Eq.~\eref{eq:gh} should be considered a function of
	$\vect{x}$, via Eq.~\eref{eq:zdef}, although for convenience
	its right-hand side has been written in terms of $\vect{z}$.
	Also, its nongaussian part has been written
	using a set of
	third-order moments $\alpha^z_{ijk}$ associated with the $z_i$.
	These satisfy $\langle z_i z_j z_k \rangle = \alpha^z_{ijk}$
	and are related to the $\alpha_{ijk}$ by the rule
	\begin{equation}
		\alpha_{ijk} = A_{il} A_{jm} A_{kn} \alpha^z_{lmn} .
	\end{equation}
	Thus, the $\alpha$ transform tensorially under the change of
	basis represented by $A_{ij}$.
	The basis functions $H_{ijk}$
	which underlie the cumulant expansion
	are products of Hermite polynomials.
	The $n^{\mathrm{th}}$ polynomial in the Hermite sequence is obtained from
	Rodrigues' formula,
	\begin{equation}
		H_n(w) = (-1)^n \e{w^2/2} \frac{\partial^n}{\partial w^n}
		\e{-w^2/2} .
	\end{equation}
	The $H_n$ satisfy an orthogonality relation,
	\begin{equation}
		\int_{-\infty}^{\infty}
			\frac{\e{-w^2/2}}{\sqrt{2\pi}}
			H_n(w) H_m(w) \; \d w = n! \delta_{mn} .
			\label{eq:ortho}
	\end{equation}
	The $H_{ijk}$ are defined by a generalized version of Rodrigues' formula.
	They satisfy
	\begin{equation}
		H_{i_1 i_2 \cdots i_n} = (-1)^n
		\exp \left( \frac{z_j z_j}{2} \right)
		\frac{\partial^n}{\partial z_{i_1} \partial z_{i_2}
		\cdots \partial z_{i_n}}
		\exp \left( - \frac{z_k z_k}{2} \right) .
		\label{eq:rodr}
	\end{equation}
	Eqs.~\eref{eq:ortho}--\eref{eq:rodr} make
	the $H_{i_1 \cdots i_n}$ orthogonal in the measure
	$\exp( - \vect{z}^2/2 ) \, \d^N z$.
	
	Eq.~\eref{eq:gh} is a cumulant expansion in the sense discussed
	in \S\ref{sec:intro}.
	It is numerically close to a Gaussian
	for small $\alpha^z_{ijk}$. It has nonzero third moments,
	which are perturbatively small, but
	all higher cumulants are zero.
	Higher $n^\mathrm{th}$ cumulants may be incorporated by
	including them in Eq.~\eref{eq:gh} as coefficients of
	$n^\mathrm{th}$ order functions $H_{i_1 \cdots i_n}$.
	Nevertheless, Eq.~\eref{eq:gh}
	is not an asymptotic expansion in the $z_i$.
	Moreover, finite truncations may be negative for some values of $z_i$.
	For our purposes the cumulant expansion is a formal tool,
	and these mild pathologies are not a cause of serious difficulty
	because
	we do not make use of the probability distribution directly.
	
	The $H_{i_1 \cdots i_n}$ obey certain important identities. By
	repeatedly commuting $z$ and $\partial/\partial z$ one can
	use the generalized form of Rodrigues' identity to
	show
	\begin{equation}
		z_m H_{i_1 \cdots i_n} =
		H_{i_1 \cdots i_n m} + \delta_{i_1 m} H_{i_2 \cdots i_n}
		+ \cdots + \delta_{i_n m} H_{i_1 \cdots i_{n-1}} .
		\label{eq:h-sum}
	\end{equation}
	Further,
	differentiating Eq.~\eref{eq:rodr} and making use of~\eref{eq:h-sum}
	one can show
	\begin{equation}
		\frac{\partial H_{i_1 \cdots i_n}}{\partial z_m}
		= \delta_{i_1 m} H_{i_2 \cdots i_n} + \cdots +
		\delta_{i_n m} H_{i_1 \cdots i_{n-1}} .
		\label{eq:h-diff}
	\end{equation}
	Eqs.~\eref{eq:h-sum} and~\eref{eq:h-diff} play a significant
	role in extracting a moment hierarchy from
	the Gauss--Hermite expansion.
	
	\subsection{Transport of the probability density}
	
	Eq.~\eref{eq:gh} is time dependent, because of the explicit
	time dependence of $\vect{X}(t)$,
	$\matrix{\Sigma}(t)$ and $\alpha_{ijk}(t)$.
	We assume that time evolution of
	$\vect{x}$ is generated by a velocity field $\vect{u}(t, \vect{x})$,
	using the rule $\dot{\vect{x}} = \vect{u}$.
	The vector $\vect{u}$	depends
	on $\vect{x}$, but may also depend explicitly
	on time. It is possible to interpret $\vect{u}$ as a
	time-dependent Hamiltonian
	vector field whose integral curves are the allowed trajectories
	in phase space.
	As time evolves,
	the shape of the probability distribution is focused and sheared
	by the action of the velocity field.
	This geometrical evolution is described by the transport
	(or ``continuity'') equation
	\begin{equation}
		\frac{\partial P}{\partial t} +
		\frac{\partial( u_i P )}{\partial x_i} = 0 .
		\label{eq:transport}
	\end{equation}
	Eq.~\eref{eq:transport} accounts for changes in shape and profile
	of $P$, but
	conserves the overall volume of the distribution.
	It is the zero-diffusion limit of a Chapman--Kolmogorov
	or Fokker--Planck equation.
	These equations occur in many areas in physics,
	including the heat equation and Schr\"{o}dinger's equation.
	
	In principle Eq.~\eref{eq:transport} could be solved directly,
	but analytic progress is possible only for a limited
	choice of $u_i$. Numerical approaches are rather involved.
	As an alternative to solving for $P$ itself,
	Eq.~\eref{eq:transport} may be converted 
	into a system of coupled equations for
	the $n^\mathrm{th}$ moments of $P$. The evolution equation for any moment
	will generally depend on all the others, but if
	increasingly high-order moments decrease in magnitude
	then it may be a reasonable approximation to truncate the coupled
	system at finite order. In what follows we will carry out
	this programme, assuming that it is only necessary to retain
	the moments $X_i(t)$, $\Sigma_{ij}(t)$ and $\alpha_{ijk}(t)$.
	
	In this section
	we are assuming that the third- and higher $n^\mathrm{th}$-order
	moments are
	perturbatively small.
	Therefore, probability is concentrated in the vicinity
	of the instantaneous centroid $X_i(t)$. The influence of the velocity
	field in reshaping the distribution
	is greatest in this region.
	Near the centroid, we find
	\begin{eqnarray}
		\fl\nonumber
		u_i = u_{i0} + u_{ij} (x_j - X_j) + \frac{1}{2} u_{ijk}
		(x_j - X_j) (x_k - X_k) + \cdots \\
		= u_{i0} + u_{ij} A_{jk} z_k + \frac{1}{2} u_{ijk}
			A_{jm} A_{kn} z_m z_n + \cdots ,
		\label{eq:vel}
	\end{eqnarray}
	where the coefficients $u_{i0}$, $u_{ij}$ and $u_{ijk}$ satisfy
	\begin{equation}
		\fl
		u_{i0}(t) = u_i(t)|_{\vect{x} = \vect{X}(t)} , \quad
		u_{ij}(t) = \left.
			\frac{\partial u_{i}(t)}{\partial x_j}
			\right|_{\vect{x} = \vect{X}(t)} , \quad \mbox{and} \quad
		u_{ijk}(t) = \left.
			\frac{\partial^2 u_{i}(t)}{\partial x_j \partial x_k}
			\right|_{\vect{x} = \vect{X}(t)} .
	\end{equation}
	A particle located at the centroid,
	$x_i = X_i(t)$, would
	evolve according to $\d x_i / \d t = u_{i0}(t)$.
	The
	presence of the higher terms in~\eref{eq:vel} reflects the way in which
	the wings of the probability distribution,
	which are absent for a point particle, sample nearby parts of the
	velocity field.  
		
	Whether or not $u_i$ depends explicitly on time, these
	coefficients will do so because they are evaluated at the
	time-dependent location $\vect{x} = \vect{X}(t)$.
	Moreover, since we must truncate Eq.~\eref{eq:vel} at finite
	order to obtain a finite system of evolution equations,
	there will be an unaccounted remainder
	which makes the effective velocity field time dependent.
	
	To proceed, we must determine $\partial P / \partial t$
	and $\partial(u_i P)/\partial x_i$.
	According to Eq.~\eref{eq:transport} their sum must be equal to zero.
	Using Eqs.~\eref{eq:h-sum} and~\eref{eq:h-diff}
	to exchange factors of $z$ or $\partial / \partial z$
	(as applied to the $H_{i_1 \cdots i_n}$)
	with sums of other $H$-functions,
	it can be cast as a Hermite tableau of the form
	\begin{equation}
		P_g \left[
			c_0 +
			c_i H_i +
			c_{ij} H_{ij} +
			c_{ijk} H_{ijk} + \cdots
		\right] = 0 .
	\end{equation}
	The orthogonality of $H_{i_1 \cdots i_n}$ in the measure
	$P_g \, \d^N z$ implies a hierarchy of equations,
	\begin{equation}
		c_0 = c_i = c_{(ij)} = c_{(ijk)} = \cdots = 0 ,
	\end{equation}
	where brackets denote symmetrization of indices.
	
	The $c_0$ equation expresses overall conservation of probability,
	and is identically satisfied because Eq.~\eref{eq:transport}
	is conservative. The equation $c_i = 0$ gives an evolution equation
	for the centroid,
	\begin{equation}
		\frac{\partial X_i}{\partial t}
		= u_{i0} + \frac{1}{2} u_{imn} \Sigma_{mn} + \cdots ,
		\label{eq:first}
	\end{equation}
	where `$\cdots$' denotes omitted terms which are higher order
	in cumulant expansion.
	The equation $c_{(ij)} = 0$ gives a similar evolution equation
	for the covariance matrix, $\Sigma_{ij}$,
	\begin{equation}
		\frac{\partial \Sigma_{ij}}{\partial t} =
		u_{im} \Sigma_{mj} + u_{jm} \Sigma_{mi}
		+ \frac{1}{2} u_{i mn} \alpha_{j mn}
		+ \frac{1}{2} u_{j mn} \alpha_{i mn}
		+ \cdots .
		\label{eq:second}
	\end{equation}
	Finally, $c_{(ijk)} = 0$ is equivalent to an evolution equation
	for the $\alpha_{ijk}$
	\begin{equation}
		\frac{\partial \alpha_{ijk}}{\partial t} =
		u_{i m} \alpha_{m jk} + u_{i mn} \Sigma_{jm} \Sigma_{kn}
		+ ( i \rightarrow j \rightarrow k ) + \cdots ,
		\label{eq:third}
	\end{equation}
	where $(i \rightarrow j \rightarrow k)$ denotes
	the preceding term with cyclic permutations
	of the indices.
	Eqs.~\eref{eq:first}--\eref{eq:third} represent the first principal
	result of this paper. They
	agree with the
	corresponding evolution
	equations for first, second and third moments obtained in
	Ref.~\cite{Mulryne:2009kh}, but apply for an arbitrary number of
	scalar fields.

	\section{Method B: Generating functions}\label{sec:cumulant}

	In this section we are going to rederive the results of
	\S\ref{sec:transport} using a technique which is valid for
	a probability distribution in the neighbourhood of an
	arbitrary kernel distribution. It can accommodate an arbitrary 
	number of fields, and applies to any order in the cumulant expansion.
	The method applied in Ref.~\cite{Mulryne:2009kh} and~\S\ref{sec:transport}
	exploited orthogonality properties of Hermite
	polynomials and was therefore restricted to a Gaussian kernel.
	Here, we take a different approach.
	We introduce generating functions for the
	cumulants, and show that this method leads to an efficient
	derivation of the evolution equations.
	
	We retain the notation of \S\ref{sec:transport}.
	In this section, we
	introduce the new method by using it to study a probability
	distribution with a single field.
	The extension to an arbitrary number of fields involves more complicated
	combinatorics, but ultimately yields expressions
	equivalent to Eqs.~\eref{eq:first}--\eref{eq:third}.
	We discuss the multiple-field case in \ref{a:cumulant}.
	
	\subsection{Generating functions}
	
	In the one-field case, we denote the single field by $x$, and assume that
	the distribution of $x$ is described by a 
	time-dependent probability distribution of the
	form $P(x,t) \, \d x$. We denote the mean value of $x$ by $X(t)$, so that
	\begin{equation}
		\label{e:defX}
		X(t) = \int x \, P(x,t) \; \d x .
	\end{equation}
	In \S\ref{sec:transport} we restricted our attention
	to the third moment, $\alpha$.
	In this section we would like to generalize the analysis to all
	orders in the moment expansion. For this reason we introduce
	moments $\mu_n(t)$, $n = 0, 1, 2, \ldots$ defined by
	\begin{equation}
		\mu_n(t) = \int \big[ x-X(t) \big]^n P(x,t) \;\d x .
	\end{equation}
	Since $P$ is properly normalized
	we must have $\mu_0 = 1$, independent of time.
	Our definition of $X$ implies
	$\mu_1 = 0$. Therefore the first nontrivial moment is the second,
	$\mu_2$. The entire infinite set of moments can be encoded
	using the ``moment
	generating function'' $M(z,t)$, defined by
	\begin{equation}
		M(z,t) \equiv \int e^{z (x-X)} P(x,t) \;\d x
		= \sum_{n=0}^\infty \frac{z^n \mu_n(t)}{n!} .
	\end{equation}
	This definition ensures
	that the $n^\mathrm{th}$ moment can be recovered using
	the identity $\partial^n_z M(0,t) = \mu_n(t)$.
	
	In \S\ref{sec:transport}
	and the foregoing discussion we have restricted our attention to
	the moments of $P$. When discussing high orders in the moment
	expansion, it is sometimes more convenient to
	make use of the cumulants of $P$ directly.
	We define the sequence of cumulants, $\kappa_n$,
	via their generating function $C(z,t)$. This satisfies
	\begin{equation}
		\label{e:defCGF}
		C(z,t) \equiv \ln M(z,t)
		= \sum_{n=0}^\infty \frac{z^n \kappa_n(t)}{n!} .
	\end{equation}
	Conversely, the moments are related to the cumulants through 
	$M(z,t) = \e{C(z,t)}$.	 Hence, knowledge of the moments is enough to
	determine all the cumulants, and vice-versa.  For example, we always
	have $\mu_1(t)=\kappa_1(t)=0$, $\mu_2(t) = \kappa_2(t)$, and
	$\mu_3(t) = \kappa_3(t)$, 
	while
	\begin{eqnarray}
	\mu_4(t)  = \kappa_4(t) + 3 \kappa_2^2(t) , \\
	\mu_5(t)  = \kappa_5(t) + 10 \kappa_2 (t) \kappa_3 (t) , \\
	\mu_6(t)  = \kappa_6(t) + 15 \kappa_2 (t) \kappa_4(t)
		+ 10 \kappa_3^2(t) + 15 \kappa_2^3(t)
	\end{eqnarray}
	and so on \emph{ad infinitum}.
	To obtain the cumulant $\kappa_n$ one need only know the
	moments up to order $n$, and vice versa.
	
	For $n > 3$
	it is the $\kappa_n$ which are most
	useful for the characterization of primordial nongaussianity.
	In the language of field theory, the cumulants are the connected correlation
	functions of $x$, while the moments are the disconnected correlation
	functions. In particular, a pure Gaussian
	distribution has only one nonzero cumulant, $\kappa_2$.
	On the other hand, all of its
	even moments $\mu_0$, $\mu_2$, $\mu_4$, \ldots,
	are nonzero. Thus, beyond third order,
	the cumulants provide a more suitable measure of departures from
	Gaussianity.
	
	\subsection{Transport equation}
	
	The transport equation is the one-dimensional version
	of Eq.~\eref{eq:transport}.
	Combining the transport equation with~\eref{e:defX} yields
	\begin{equation}
		\label{e:tmp1}
		\frac{\d X}{\d t} = \int x \frac{\partial }{\partial t} P(x,t) \;\d x
		= \int	u(x) P(x,t)	 \; \d x
	\end{equation}	
	As in \S\ref{sec:transport},
	we assume that $x$ evolves under the influence of a velocity
	field which can be expanded around the instantaneous centroid
	$X(t)$,
	\begin{equation}
		\label{e:defux}
		u(x) = \sum_{n=0}^\infty \frac{u_n(t)}{n!} [ x-X(t) ]^n ,
	\end{equation}
	where,
	in comparison with Eq.~\eref{eq:vel},
	we have adopted a slightly different notation
	in which $u_n$ denotes the $n^{\mathrm{th}}$ derivative
	$\d^n u / \d x^n$.
	Returning to~\eref{e:tmp1} and applying~\eref{e:defux} 
	together with the definition of the moments, we
	find
	\begin{equation}
		\label{e:dotX}
		\frac{\d X}{\d t} = \sum_{n=0}^\infty \frac{\mu_n(t) u_n(t)}{n!}
		= u_0(t) + \frac{1}{2} \mu_2(t) u_2(t) + \cdots
	\end{equation}
	
	With the evolution of $X(t)$ in hand, we can derive the evolution of 
	all other cumulants.  Using~\eref{e:defCGF}, we conclude
	that the time derivatives $\d\kappa_n / \d t$ obey
	\begin{equation}
		\label{e:ctdGF}
		\sum_{n=0}^\infty \frac{z^n}{n!} \frac{\d \kappa_n(t)}{\d t}
		=
		\frac{1}{M(z,t)}
		\sum_{n=0}^\infty \frac{z^n}{n!} \frac{\d \mu_n(t)}{\d t} ,
	\end{equation}
	from which it follows that
	$\d C/\d t$ is the generating function for the cumulant time
	derivatives. Following steps similar to
	those which led us to~\eref{e:tmp1}, it can be seen that
	the evolution equation for each moment can be written
	\begin{equation}
		\label{e:mueom}
		\frac{\d \mu_n(t)}{\d t}
		= \sum_{k=0}^\infty \frac{n}{k!}
		\Big[
			\mu_{n+k-1}(t) - \mu_{n-1}(t)\mu_k(t)
		\Big]
		u_k(t)
	\end{equation}
	Inserting this expression in~\eref{e:ctdGF} yields a
	generating function
	for the time derivatives of each cumulant.
	The result can be written in terms of the moments.
	Finally, using the relationship between the moments and cumulants,
	the cumulant evolution equations can be written in terms of the
	cumulants alone. This gives a closed system of equations for the 
	cumulants, with an infinite number of variables.
	
	In practice, only a finite number of variables can be evolved and
	it is necessary to truncate both the series of cumulants
	and the expansion of $u$.
	If the Taylor expansion of the velocity field is truncated,
	then it is evident from~\eref{e:mueom} that the evolution
	equation for each moment involves moments of higher order.
	Accordingly, the time-evolution of an individual cumulant always
	involves cumulants of higher order. Thus, even when the velocity field
	is truncated, the system of cumulant evolution equations
	does not close at any finite order. We must
	choose an order at which to approximate the full cumulant expansion.
	If desired, we can obtain an estimate of the error involved in
	truncation at this order by determining the degree to
	which time evolution sources higher
	cumulants, forcing them to become nonzero.

	We now illustrate this procedure in operation using a simple example.	
	Suppose we wish to carry the velocity
	field expansion to
	third order in $(x-X)$, so that
	\begin{equation}
		\fl
		u(x) = u_0(t) + [ x-X(t) ] u_1(t)
		+ \frac{1}{2!} [x-X(t)]^2 u_2(t) + \frac{1}{3!} [x-X(t)]^3 u_3(t) .
	\end{equation}
	Assume further that the maximal nonzero cumulant is of order four.
	Applying the formulae above, we find
	\be
		\frac{\d X}{\d t} =
		u_0(t) + \frac{1}{2!} u_2(t) \kappa_2(t)
		+ \frac{1}{3!} u_3(t) \kappa_3(t)
		+ \frac{1}{4!} \Big[ 
		3 \kappa_2(t)^2 + \kappa_4(t) \Big] .
	\ee
	The cumulant evolution equations enforce $\d \kappa_0 / \d t = 0$ and
	$\d \kappa_1 / \d t = 0$.	Furthermore,
	the cumulants $\kappa_2$, $\kappa_3$ and $\kappa_4$
	evolve according to
	\begin{eqnarray}
		\label{eq:kappa2}
		\frac{\d \kappa_2}{\d t} =
		2 u_1 \kappa_2 + u_2 \kappa_3 + u_3 \Big[	 \kappa_2^2 
		+ \frac{1}{3}	 \kappa_4 \Big] \\
		\label{eq:kappa3}
		\frac{\d \kappa_3}{\d t} = 3 u_1 \kappa_3 
		+ u_2 \Big[ 3 \kappa_2^2 + \frac{3}{2} \kappa_4 \Big] 
		+ \frac{9}{2} u_3	\kappa_2 \kappa_3	 \\
		\label{eq:kappa4}
		\frac{\d \kappa_4}{\d t} =	4 u_1 \kappa_4
		+ u_2 \Big[ 12 \kappa_2 \kappa_3 + 2 \kappa_5 \Big]
		+ u_3 \Big[ 4 \kappa_2^3 + 6 \kappa_3^2 + 8 \kappa_2 \kappa_4 \Big] ,
	\end{eqnarray}
	where we have suppressed the time dependence of the $\kappa_j$ and 
	$u_j$.

	Under our assumptions, the evolution equation for $\kappa_5$ is
	\begin{equation}
		\frac{\d \kappa_5}{\d t} =
		 u_2 \Big[ 20 \kappa_2 \kappa_4 + 15 \kappa_3^2 \Big]
		 + u_3 \Big[ 30 \kappa_2^2 \kappa_3 + 25 \kappa_3 \kappa_4 
		 \Big]
		 \label{eq:four-error}
	\end{equation}
	If a truncation to fourth-order quantities were self-consistent,
	this equation should vanish
	as a consequence of our assumption
	that $\kappa_n = 0$ for $n \geq 5$.
	That it does \emph{not} vanish is a measure of the error involved
	in our truncation.
	As explained above, a similar effect occurs no matter at which order
	the truncation is made.
	In the analysis of this section, we have made no assumption that
	either the moments or cumulants are small, or order themselves
	into an ultimately decreasing sequence.
	Eq.~\eref{eq:four-error} demonstrates that the truncated hierarchy
	will be a useful predictive instrument only when a negligible
	variation in $\kappa_5$ is sourced over the time interval of interest.
	This will typically (but not absolutely) require the cumulants
	to fall in an ordered structure, for example
	$|\kappa_2| > |\kappa_3| > |\kappa_4|$.

	For example, we may suppose that $|\kappa_n| = \Or(\delta^n)$
	where $\delta \ll 1$ is a small positive number.
	Eq.~\eref{eq:four-error} shows that
	source term for $\kappa_5$ is of order $\Or(\delta^6)$,
	and therefore after a short time interval $\Delta t$
	we can expect $\kappa_5 \sim (\Delta t) \delta^6$.
	Truncation to
	fourth order, setting $\kappa_5$ and all higher cumulants to zero,
	may be acceptable approximation over sufficiently short times
	that $\kappa_5$ does not grow to the degree that it contaminates
	any observable of interest. How long this time can be is model
	dependent.
	A similar analysis can be given for all higher cumulants.
	Growing secular terms which eventually invalidate
	perturbation theory are a typical feature in studying the evolution
	of fluctuations. They are encountered directly in the
	Feynman amplitudes of quantum field theory, and in the
	most common implementations of the separate universe picture.

	Eqs.~\eref{eq:kappa2}--\eref{eq:kappa4}
	coincide with Eqs.~\eref{eq:second} and~\eref{eq:third}
	when specialized to the single-field case,
	but include the contribution of the third derivative
	of the velocity field, $u_3$.
	Eq.~\eref{eq:kappa4} gives, for the first time, the evolution
	of the kurtosis in a single-field setting. 
	In~\ref{a:cumulant}
	we apply this method to the case of multiple fields,
	and again 
	find agreement with the results of \S\ref{subsec:gauss}.
	
	\section{Inflationary perturbations}
	\label{sec:inf}
	
	We now wish to apply the general framework assembled in
	\S\S\ref{sec:transport}--\ref{sec:cumulant} to
	the fluctuations which
	are generated and subsequently evolve during an inflationary era.
	To do this we identify the variables $x_i$ of
	\S\S\ref{sec:transport}--\ref{sec:cumulant}
	with the light, scalar degrees of freedom which are
	excited at that time.
	We must also identify a choice of time variable, labelled
	$t$ in \S\S\ref{sec:transport}--\ref{sec:cumulant},
	which corresponds to a choice of slicing in the spacetime picture.
	
	\subsection{Gauge choices}
	\label{sec:gauge}
	
	Consider a theory of $\N$ scalar fields coupled to a metric
	theory of gravity.
	The degrees of freedom in this system are the $\N$ fields
	themselves, $\phi_i$ for $i \in \{ 1, \ldots, \N \}$,
	together with the lapse and shift functions,
	written $N$ and $N^a$,
	and the spatial 3-metric $h_{ab}$,
	\begin{equation}
		\d s^2 = - N^2 \, \d t^2 + h_{ab} ( \d x^a + N^a \, \d t)
		( \d x^b + N^b \, \d t ) .
		\label{eq:adm}
	\end{equation}
	In Einstein gravity
	the lapse and shift are not propagating fields, and are eliminated
	by constraint equations. The 3-metric $h_{ab}$
	encodes six independent degrees of freedom but
	in this paper we concentrate on the two spin-0 modes,
	of which only one is physical. We take this to be
	the volume modulus $\det h_{ab}$;
	the other can be absorbed by a spatial coordinate redefinition.
	We write $\det h_{ab} = \e{6(N - N_0)} \delta_{ab}$, and refer to $N$
	as the \emph{integrated e-foldings of expansion}.
	The zero point $N_0$ is arbitrary.
	Note that the integrated number of e-folds is quite separate from the
	lapse function which occurs in Eq.~\eref{eq:adm}, also traditionally
	denoted $N$. In what follows we shall
	work in an unperturbed universe for which
	the lapse is unity. Therefore we have no need to refer to it
	explicitly, so that the appearance of $N$
	without qualification is
	unambiguous.
	
	We now have $M+1$ variables: the $M$ fields, and the integrated
	expansion $N$.
	Not all these are independent, and
	one of them can be eliminated by a
	suitable choice of time, $t$. Whatever choice we make, surfaces of
	constant $t$ foliate spacetime into spatial hypersurfaces
	which we refer to as a slicing.
	The transport equation, Eq.~\eref{eq:transport}, evolves
	a probability distribution for the $x_i$ from one slice in this
	foliation to the next.
	There are two slicings of particular importance for inflationary
	fluctuations.
	
	\para{Spatially flat slicing.}
		In this gauge, spatial hypersurfaces are chosen
		so that the integrated number of e-foldings,
		$N$, is uniform across the slice.
		The scalar fields $\phi_i$ fluctuate from place to place.
		To apply
		the formalism of \S\ref{sec:transport}
		we work on slices of uniform expansion $N$, and
		smooth the scalar fields on a comoving lengthscale $L$ to
		give an ensemble of $L$-sized regions.

		These regions traverse a bundle of adjacent trajectories in field
		space, with some characteristic dispersion and higher-order
		moments which it is our intention to calculate.
		To make contact with observation the bundle should be chosen
		so that every trajectory reheats almost surely in our local vacuum.
		With this choice
		we can suppose the dispersion and higher-order moments
		of $L$-sized regions in our observable universe should be similar
		to those of the bundle. On scales comparable to the size of
		our observable patch we sample only
		a small number of trajectories, making the prospect of a mismatch
		with the bundle average (``cosmic variance'')
		more likely.
		
		The centroid of the bundle follows a path
		$\Phi_i^L(t)$ in field space, where the superscript `$L$'
		indicates dependence on the smoothing scale.%
			\footnote{Note that $\Phi_i^L(t)$ need not be an integral curve
			of the velocity field, and therefore may not constitute an allowed
			inflationary trajectory.}
		The probability distribution we wish to calculate
		is a function of the scalar fields,
		$\phi_i = \Phi_i^L(t) + \delta \phi_i^L$.
		We take
		$x_i = \phi_i / \Mp$ and $X_i(t) = \Phi_i^L(t) / \Mp$.
		The probability density obtained in this way
		gives information about the distribution of field values on the
		scale $L$ only. To obtain a relation
		between bundles smoothed on different scales $L$ and $L'$,
		two separate distributions must be computed and their information
		combined. This would be necessary, for example, to obtain the
		spectral index.

		We also require initial conditions, set
		at the time $t_L$
		when the wavenumber corresponding to the smoothing scale $L$
		crosses the horizon.
		We pick initial expectation values
		$\Phi_i^L(t_L)$ centred on the inflationary trajectory of interest.
		In the spatially flat
		slicing, the joint probability distribution of fluctuations in the
		scalar fields at time $t_L$ can be calculated directly
		\cite{Bardeen:1983qw,Guth:1982ec,Hawking:1982cz,Hawking:1982my,
		Mukhanov:1985rz,Sasaki:1986hm},
		with each field acquiring
		a variance of order $\langle \delta \phi_i^2 / \Mp \rangle_L
		= \sigma^2_L$,
		where $\sigma_L = H_L / \Mp \sim 10^{-5}$
		and $H_L$ is the Hubble rate
		when the wavenumber $1/L$ crosses the horizon.
		The skewness of the bundle is negligible, with
		$\alpha_{ijk}^L \sim \sigma_L^4 \approx 0$
		\cite{Maldacena:2002vr,Seery:2005gb}.
		In addition, to leading order in
		$\epsilon = - \dot{H}/H^2$,
		the slow-roll condition makes the fields uncorrelated
		at horizon exit,
		with $\langle \delta \phi_i \delta \phi_j / \Mp^2 \rangle_L
		\sim \epsilon_L\sigma^2_L$ if $i \neq j$.%
			\footnote{The conclusion that the $\delta \phi_i$
			are virtually uncorrelated at horizon exit
			implies that the
			inflationary trajectories---described by the integral
			curves of the velocity field $u_i$ in this gauge---are
			effectively straight lines
			for a few e-folds around horizon-crossing,
			up to corrections of $\Or(\epsilon)$.
			This applies in canonical inflationary models,
			but in more general examples this may not occur
			\cite{Chen:2009zp}.
			In such cases the theory becomes more complicated and
			the formalism of this paper will no longer apply.}
		Taking the initial
		dispersion to be given by $\sigma_L$ and setting
		$\alpha^L_{ijk}$ and any cross-correlations to
		be initially zero, we can evolve the $\sigma$ and $\alpha$ 
		from horizon crossing until any desired future time, such as 
		the end of inflation.	 At this point the field's moments on 
		the flat hypersurface can be used to calculate the moments 
		of $\zeta$ using a gauge transformation.	For two fields this 
		proceedure was performed in Ref.~\cite{Mulryne:2009kh}, and the 
		technology developed in this paper makes this possible 
		for an arbitrary number of fields.
	
	\para{Uniform density slicing.}
		Alternatively, we may choose our spatial hypersurfaces so that
		the Hubble rate, $H$, is uniform over the slice.
		In Einstein gravity the Friedmann constraint enforces
		$3 \Mp^2 H^2 = \rho$, so slices of uniform $H$ are also slices of
		uniform density.
		On this slicing the integrated number of e-foldings, $N$,
		will typically vary from place to place.
		The counting of independent fluctuations is the same
		as in the spatially flat slicing, because the Friedmann constraint
		makes \emph{one} field a function of all the others
		and the Hubble rate, $H$.
		Without loss of generality we can suppose that
		$\phi_{\N} = \phi_{\N}( \phi_1, \ldots, \phi_{\N - 1}, H)$.
		For notational convenience we define
		$\iso_i = \phi_i / \Mp$.
		To apply the framework of
		\S\S\ref{sec:transport}--\ref{sec:cumulant}
		we must set the time variable,
		$t$, to equal $H/\Mp$ and choose the variables
		$x_i$ whose distribution
		we wish to calculate to be
		$\{ \iso_1, \ldots, \iso_{\N - 1}, N \}$.
		
		It is still necessary to choose a smoothing scale, $L$.
		The centroid of the bundle is characterized by the expectation values
		of the first $\N -1$ scalar fields,
		$\{ \Phi_1^L(t), \ldots, \Phi_{\N - 1}^L(t) \}$
		together with the mean integrated expansion experienced by
		trajectories within the bundle, $\bar{N}_L(t)$.
		We write $N = \bar{N}_L(t) + \zeta_L$, where $\zeta_L$
		is the uniform density gauge curvature perturbation
		smoothed on scale $L$.
		On the large scales we are considering,
		the uniform density gauge and comoving gauge
		coincide which makes $\zeta_L$ numerically equal to the comoving
		curvature perturbation, $\mathcal{R}_L$.
	
	\subsection{Moment transport in the uniform density slicing}
	\label{sec:uniform}
	
	To implement moment transport in the uniform density slicing,
	we will require initial conditions for the dispersion and higher
	moments of the fluctuations
	$\{ \iso_1, \ldots, \iso_{\N-1}, \zeta \}$.
	These have not yet been calculated directly, 
	but can be obtained from
	the joint probability distribution in the spatially flat slicing
	\cite{Seery:2005gb} in conjunction with gauge transformations relating 
	super-horizon quantities in the uniform density gauge to 
	those in spatially flat 
	gauge. These transformations can be obtained using coventional
	cosmological perturbation theory \cite{Malik:2008im},
	or using the separate universe assumption,
	which was the approach taken in Ref.~\cite{Mulryne:2009kh}.
	Using the separate universe assumption, one writes
	$\zeta$ using the `$\delta N$' formula
	\cite{Lyth:2005fi}
	\begin{equation}
		\zeta(t) = N( t, \phi_\ast + \delta \phi_\ast ) - N(t, \phi_\ast )
		= N_{,i} \delta \phi_{i \ast} +
		\frac{1}{2} N_{,ij} \delta \phi_{i \ast} \delta \phi_{j \ast} + \cdots
		,
		\label{eq:dN}
	\end{equation}
	where $N(t, \phi_\ast)$ measures the
	e-foldings between
	a spatially flat slice on which the fields take
	prescribed values $\phi_{i \ast}$
	and a subsequent uniform density slice at time $t$.
	We are considering
	the special case of flat and uniform density slices which
	coincide on average, and are therefore only perturbatively separated.
	The coefficients satisfy $N_{,i} = \partial N / \partial \phi_{i \ast}$
	with similar definitions for $N_{,ij}$ and higher derivatives.

	\para{Initial conditions for isocurvature fields.}
	The fluctuations $\{ \iso_1,
	\ldots, \iso_{\N -1} \}$ can be calculated using an analogous
	formula, as described in Ref.~\cite{Mulryne:2009kh},
	\begin{equation}
		\iso_i \Mp = \frac{\partial \phi_i^c}{\partial \phi_{j \ast}}
			\delta \phi_{j \ast} +
			\frac{1}{2}
			\frac{\partial^2 \phi_i^c}{\partial \phi_{j \ast}
				\partial\phi_{k \ast}}
			\delta \phi_{j \ast} \delta \phi_{k \ast} +
			\cdots ,
		\label{eq:ys}
	\end{equation}
	The superscript `$c$' denotes scalar fields evaluated on a
	comoving (uniform density) spatial slice, in the same way that
	`$\ast$' denotes fields evaluated on spatially flat slices.
	Eq.~\eref{eq:ys} gives each field $\iso_i$ a dispersion $\sigma_i^L$,
	approximately satisfying
	\begin{equation}
		\sigma_i^L \approx
		\frac{1}{4\pi} \frac{H_L^2}{\Mp^2}
		\sum_j
			\frac{\partial \phi^c_i}{\partial \phi_{j \ast}}
			\frac{\partial \phi^c_i}{\partial \phi_{j \ast}} ,
		\hfill
		\mbox{(no sum on $i$)}
		\hspace{1cm}
	\end{equation}
	together with negligible third moments and cross-correlations.
	By a suitable choice of origin we can arrange that $N = 0$ on the
	initial spatially flat slice, and therefore $\bar{N}$ satisfies
	\begin{equation}
		\bar{N}_L = \frac{1}{2} N_{,ij}
		\langle \delta \phi_{i \ast} \delta \phi_{j \ast} \rangle_L
		= \frac{1}{8 \pi^2} \frac{H_L^2}{\Mp^2} \sum_{i} \Mp^2 N_{,ii}
	\end{equation}
	on the initial uniform density slice.
	Since this is very small it is a reasonable approximation
	to take $\bar{N}_L \approx 0$.
	Eq.~\eref{eq:dN} generates nonzero correlations between $\zeta$
	and the $\iso_i$,
	allowing us to calculate the covariances
	$\langle \zeta \iso_i \rangle_L$.
	It can also be used to obtain
	the $\zeta$-dispersion,
	$\langle \zeta \zeta \rangle_L$. We find
	\begin{eqnarray}
		\langle \zeta \iso_i \rangle_L
		\approx \frac{\Mp}{4\pi^2}\frac{H_L^2}{\Mp^2}
			\sum_i N_{,i} \frac{\partial \phi^c}{\partial \phi_{i\ast}} \\
		\langle \zeta \zeta \rangle_L
		\approx \frac{1}{4\pi^2} \frac{H_L^2}{\Mp^2} 
			\Mp^2 \sum_i N_{,i} N_{,i} .
	\end{eqnarray}
	
	\para{Expressions for velocity field.}
	With this choice of variables and
	the assumption that third-order moments are zero initially%
		\footnote{It is clearly possible to extend the approach
		we have
		described to calculate the initial third order moments
		in the uniform density	
		gauge, given the known initial conditions in the flat gauge.	
		This requires the use of $N_{,i j}$ and
		$\partial^2 \phi^{c}_i/\partial \phi_{j k *}$. The 
		assumption that these are 	
		zero, however, introduces an error only of the same order as
		that already present from
		the typical approximation of taking the initial third moments
		to be zero in the flat gauge.},
	the initial conditions, Eqs.~\eref{eq:first}--\eref{eq:third},
	can be used to compute
	the covariance matrix and third-order moments
	at any later time, provided expressions can be found for the
	velocity potential and its derivatives. 
	In the uniform density gauge, the allowed trajectories are
	integral curves of
	\begin{equation}
		u_i = \Mp \frac{\d \iso_i}{\d H}
		\quad
		\mbox{and}
		\quad
		u_N = \Mp \frac{\d N}{\d H} .
	\end{equation}
	Applying the slow-roll approximation allows us to write the velocity
	potential as a function 
	of $\iso_i$ and $H$, and we find
	\begin{equation}
		u_i = - \frac{\Mp}{H}
			\frac{\sqrt{2 \epsilon_i}}{\epsilon}
		\label{eq:ui}
	\end{equation}
	and
	\begin{equation}
		u_N = - \frac{\Mp}{H \epsilon} .
		\label{eq:uN}
	\end{equation}
	In order to write these
	formulae
	we have defined a set of partial slow-roll parameters, $\epsilon_i$,
	which satisfy
	\begin{equation}
		\epsilon_i \equiv \frac{1}{2 \Mp^2} \frac{\dot{\phi}_i^2}{H^2} ,
	\end{equation}
	where $\dot{\phi}=\partial V / \partial \phi_i / 3 H$.
	To leading order in the slow-roll approximation,
	$\epsilon = \sum_i \epsilon_i$.
	Note that both $u_i$ and $u_N$ are independent of $N$, but depend
	explicitly on the time variable $H$.
	Therefore, all $N$-derivatives of the velocity field vanish identically.
	Derivatives of Eqs.~\eref{eq:ui} and~\eref{eq:uN} with respect
	to the $\iso_i$ can be found after using the Friedmann constraint
	to eliminate the
	variations $\partial \phi_{\N} / \partial \phi_i$
	and $\partial \phi_{\N} / \partial H$.
	
	\subsection{Numerical performance of the moment transport method}

	The imminent arrival of high-quality microwave background data
	from the \emph{Planck} satellite implies that it will soon be
	necessary to obtain accurate estimates of the nonlinearity parameters
	in a wide range of inflationary models.
	Although analytic formulas for $\fnl$ exist, they
	are available only for certain forms of the potential
	and even when their
	use is possible they rapidly become unwieldy
	in the limit of a large number of fields.
	For these reasons we expect numerical methods for computing
	$\fnl$ to become of increasing importance.
	In \S\ref{sec:numerics} below,
	we discuss a numerical implementation of the moment transport
	method using the uniform density slicing.
	Here, we briefly comment on the
	computational efficiency of the moment transport
	algorithm in comparison with alternative approaches.
			
	In a system with $M$ fields, the moment transport method requires
	a solution of Eqs.~\eref{eq:first}--\eref{eq:third}. These comprise	 
	$M$ unique equations for the centroid, 
	$M(M + 1)/2$ for the covariance matrix and
	$M(M + 1)(M+2)/6$ for the 3-point functions.
	Therefore, for large $M$ we must solve a coupled system of
	$\Or(M^3)$ equations.
	In comparison with popular alternative methods based on
	the $\delta N$ formula
	\cite{Starobinsky:1982ee,Starobinsky:1986fxa,Sasaki:1995aw,Lyth:2005fi},
	we argue that the transport method is computationally
	simpler.

	To make use of the $\delta N$ formula requires calculation of
	the derivatives
	$\partial N / \partial \phi_{i \ast}$,
	$\partial^2 N / \partial \phi_{i \ast} \partial \phi_{j \ast}$,
	\ldots, and so on.
	Therefore
	a direct implementation of the $\delta N$ formula with $M$ fields
	requires numerical evolution of the background field equations
	for many initial conditions,
	from which the necessary derivatives may be extracted.
	
	How many evolutions of the background equations are required? 
	This will determine the computational efficiency.
	To calculate $\fnl$, we require derivatives up to second order but not
	higher.
	The first-order derivatives may be obtained by taking finite differences
	between inflationary trajectories with initial conditions separated
	by a small distance $\delta$, giving derivatives up to an error
	of order $\delta^2$.
	This requires
	of order $M+1$ evolutions of the $M$ equations, or
	the solution to $\sim M^2$ ordinary differential equations.%
		\footnote{In practice, a more accurate discretization scheme may be
		required.}
	Extending this argument to the 
	second order derivatives shows that to determine $\fnl$
	we must solve
	$\sim M^3$ ordinary differential equations.
	For higher moments the same counting applies,
	so that an evaluation of the trispectrum
	will typically
	require the solution to $\sim M^4$
	ordinary differential equations, or more generally $\sim M^n$ for the
	amplitude of $n$-point correlations.
	
	We conclude that, to compute $\fnl$, both the transport method and a direct
	$\delta N$ require the solution of $\Or( M^3 )$
	ordinary differential equations.
	Asymptotically, their relative efficiency depends on details of the
	algorithm.
	However,
	the method of moment transport is especially simple.
	Numerical
	$\delta N$
	requires a discretization scheme to compute the partial
	derivatives.
	The final accuracy can depend on our choice of discretization.
	Also, as we will discuss below, $\delta N$ requires
	the e-folds of expansion to be determined very accurately on
	successive time slices.
	In comparison, 
	the transport method 
	requires only the solution to a set of coupled
	ordinary differential equations.  This allows
	off-the-shelf differential equation solvers to be brought to bear 
	on the problem immediately.
	
	In a direct $\delta N$ algorithm,
	very high accuracy is required when evolving the background field
	equations
	because at the end of the calculation we must take
	finite differences to construct the derivatives of $N$.
	In simulations with up to $M=5$ fields we have found this process
	to be
	sensitive to small
	numerical inaccuracies;
	the moment transport algorithm produces
	accurate results with larger numerical tolerances.
	This observation can be explained in a simple way. Suppose we 
	wish to compute $\delta N$ to fractional accuracy $f$, where
	\begin{equation}
		\label{e:TargP}
		f = \frac{\mbox{Error}(\delta N)}{\delta N}
	\end{equation}
	On the one hand, if we use the na\"{\i}ve $\delta N$ algorithm and
	compute $N$ using
	an integration routine
	which operates at fractional accuracy $f_1$, then the
	absolute error in $N$ will be $f_1 N$.
	Therefore, the absolute error in
	$\delta N$ is also $f_1 N$, and to achieve the target
	precision~\eref{e:TargP} we must
	choose $f_1 \sim f \delta N/N$.
	On the other hand, the moment transport approach essentially
	integrates $\delta N$ directly. Using an integration
	routine with fractional
	accuracy $f_2$ we will evaluate $\delta N$ with the same fractional
	accuracy, so $f_2 \sim f$.
	Since $\delta N/N \ll 1$, we have $f_1 \ll f_2$.
	We conclude that the moment transport
	algorithm can operate with much lower numerical tolerances than the
	na\"{\i}ve $\delta N$ approach.

	Finally, we note that direct implementation of $\delta N$ is not the
	only possibility.
	Yokoyama, Suyama \& Tanaka
	\cite{Yokoyama:2007uu,Yokoyama:2007dw}
	suggested an approach which is broadly similar to that employed in
	Ref.~\cite{Mulryne:2009kh},
	reviewed in Ref.~\cite{Tanaka:2010km}.
	In this approach, one seeks to calculate the field perturbations on
	a uniform curvature hypersurface at the time of interest, as
	functions of their initial values, using the $\delta N$ formula
	to effect the final gauge transformation.
	One important difference between the formulation of this paper
	(and Ref.~\cite{Mulryne:2009kh}) and that of Yokoyama {\etal}
	is that the authors of Refs.~\cite{Yokoyama:2007uu,Yokoyama:2007dw}
	continue to work in terms of the field perturbations, rather
	than evolving the moments of the distribution.
	We believe this makes our formulation
	simpler to implement in practice.

	\section{Numerical examples}
	\label{sec:numerics}
	
	In this section we present results
	obtained from the moment transport method,
	implemented on a desktop computer. This demonstrates that it is possible
	to compute $\fnl$ in systems
	with at least $M \sim 10^2$ fields using commodity 
	hardware.
	The exact time required for a simulation
	is highly model dependent, but a simple $M=10$
	example may only
	take a few seconds, while a 
	$M=100$ calculation can be carried out in less than an hour.
	Our numerical code is implemented using Matlab, and evolves
	the uniform density gauge
	transport equations,
	Eqs.~\eref{eq:first}--\eref{eq:third}.
	Using this code we are able to obtain
	numerical solutions for the potential
	\cite{Dimopoulos:2005ac,Battefeld:2006sz,Kim:2006ys,Kim:2006te}
	\begin{equation}
		V= \sum_i \frac{1}{2} m_i^2 \phi_i^2 .
		\label{eq:nflation-potential}
	\end{equation}
	This describes a number of uncoupled fields with quadratic potentials.
	It is also
	the small-field approximation to a collection of uncoupled
	axions, which have trigonometric potentials;
	this latter case is often known as Nflation.
	The variables evolved in this gauge and their initial conditions
	are determined by applying the discussion of \S\ref{sec:uniform}.
	In the special case where all $m_i$ are equal
	there is an $O(M)$ symmetry, and the fields roll radially to the
	origin. Where this symmetry exists we have verified that our code
	reproduces the expected single-field result of constant
	$\zeta$ and negligible $\fnl$, for up to $10^2$ fields.

\begin{figure}
		\center{\includegraphics[width = 10cm]{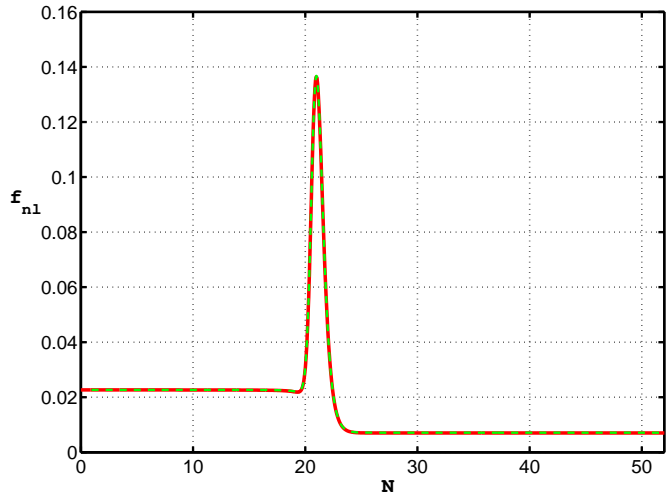}}
		\caption{Evolution of $\fnl$ (solid red line)
 			calculated using the moment transport equations 
			for a two-field Nflation model, working in the quadratic
			approximation.
			The green dashed line shows
			$\fnl$ calculated for 
			the same model using the $\delta N$ formula.
			Initial conditions are described in the text.
			In this and subsequent figures the horizon axis is labelled
			by $N$, the number of e-folds which have elapsed since
			horizon-crossing of the wavenumber of interest.
			\label{fig1}}
	\end{figure}

\begin{figure}
		\center{\includegraphics[width = 10cm]{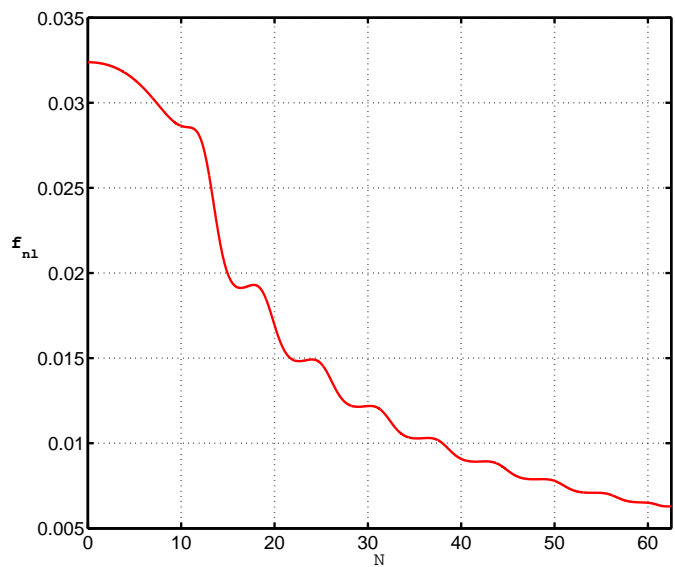}}
		\caption{Evolution of $\fnl$
 			calculated using the moment transport equations 
			for a $10$-field Nflation model
			in the quadratic approximation.
			The initial conditions and distribution of masses
			are described in the main text.
			\label{fig2}}
	\end{figure}

\begin{figure}
		\center{\includegraphics[width = 10cm]{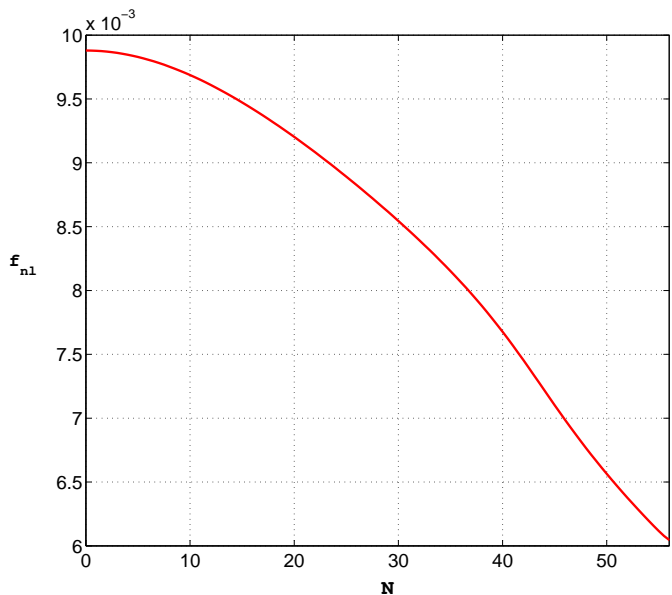}}
		\caption{Evolution of $\fnl$
 			calculated using the moment transport equations 
			for a $100$-field Nflation model
			in the quadratic approximation.
			The initial conditions and distribution of masses are
			described in the main text.
			\label{fig3}}
	\end{figure}

	Where the $m_i$ are different,
	it is known from analytic calculation that this potential does not 
	give rise to a large nongaussianity \cite{Vernizzi:2006ve,Alabidi:2005qi}.
	Our results confirm this conclusion, but the simplicity of the
	model and the existence of analytic predictions
	makes it a useful test of our method.
	In Figs.~\ref{fig1}--\ref{fig3} we show illustrative results.
	Consider first the case of a small number of fields.
	Rigopoulos, Shellard \& van Tent \cite{Rigopoulos:2005ae,Rigopoulos:2005us}
	analysed a two-field example with mass ratio $m_1 = 9 m_2$.
	Vernizzi \& Wands later studied the same model using a combination of
	analytic and numerical methods \cite{Vernizzi:2006ve}
	and concluded that
	$\fnl$ at the end of inflation was very small,
	of order $10^{-2}$.

	We depict the evolution of $\fnl$ in Fig.~\ref{fig1}. For comparison,
	we plot a calculation of $\fnl$, for the same model and
	initial conditions,
	using a
	numerical slow-roll implementation of the $\delta N$ formula,
	finding complete agreement.
	In this and subsequent figures the horizontal axis is labelled by
	$N$, the number of e-folds which have elapsed since
	horizon-crossing of the wavenumber of interest.
	Because our moment-transport and $\delta N$ calculation are performed
	using the
	slow-roll equations of motion, we do not need to allow a
	`lead time' for the simulation to converge to the slow-roll
	attractor.
 
	The qualitative behaviour of $\fnl$ in this model was discussed
	at the end of \S\ref{sec:uniform}.
	Our evolution agrees with this discussion,
	and also the explicit calculations of Vernizzi \& Wands
	\cite{Vernizzi:2006ve} and
	the moment transport method using the spatially flat slicing
	\cite{Mulryne:2009kh}.
	The most prominent feature is a well-documented
	spike, which occurs when the heavier field reaches the vicinity of
	its minimum and decouples from the dynamics.

\begin{figure}
		\center{\includegraphics[width = 10cm]{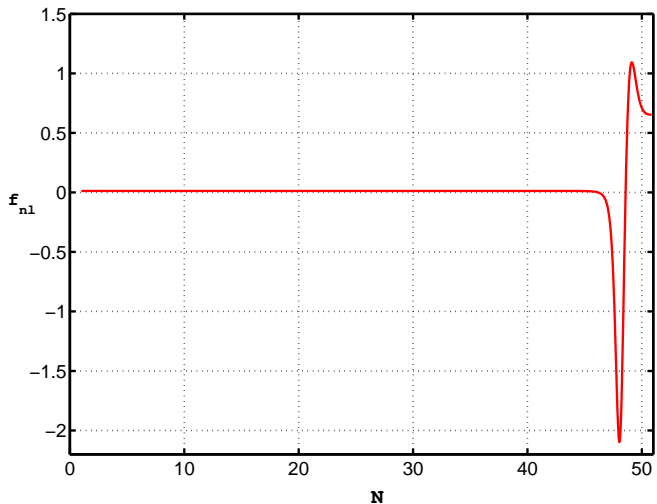}}
		\caption{Evolution of $\fnl$ 
 			calculated using the moment transport equations 
			for a $100$-field Nflation model,
			retaining the full trigonometric form of the potential.
			The initial conditions are chosen so that a single field
			explores the region close to the hilltop. The
			nongaussianity of this field dominates the late-time
			attractor solution, at which $\fnl$ converges to
			a time-independent nonzero value.\label{fig4}}
	\end{figure}

	In Fig.~\ref{fig2},  we increase the number of fields to $10$.
	For simplicity we
	take $\phi_i =5$ for 
	all fields and distribute the masses
	logarithmically, in such a way that $m_i= 2  m_{i-1}$.
	As successive fields evolve to their minima, a wavelike structure is
	produced in $\fnl$.
	Its value at the end of inflation is again of order $10^{-2}$.
	Kim \& Liddle \cite{Kim:2006te}
	argued that, at the end of inflation in a model
	of the form~\eref{eq:nflation-potential}, the
	bispectrum imprinted
	for a mode of wavenumber $k$
	could be parametrized by
	$(6/5) \fnl \approx 1/2N_\ast$.%
		\footnote{We have neglected a quantum-mechanical contribution
		generated by interference among field modes at horizon crossing.
		This contribution is not determined by the moment transport
		method, and its contribution is not represented in
		Figs.~\ref{fig1}--\ref{fig4}. It is known to be small
		\cite{Lyth:2005qj,Vernizzi:2006ve}
		and may be neglected when $|\fnl| \gtrsim 1$.}
	In this estimate, $N_\ast$ is the number of e-folds
	to the end of inflation from the field values at horizon exit of
	mode $k$, dropping any
	correction from the end of inflation.
	Within this approximation,
	the result is independent of the masses,
	initial conditions, or number of
	fields.
	In this model, we find
	$N_\ast \approx 63$ yielding
	a value for $\fnl$ in good agreement with
	the formula of Kim \& Liddle.

	In order to demonstrate the ability of our algorithm to deal with a
	large number of fields, 
	we give two examples with $M = 10^2$ fields.
	First, we retain the potential~\eref{eq:nflation-potential}
	and set $\phi_i = 1.5 \Mp$ for each field.
	We distribute masses such that $m_i = m_{i-1} + 0.1 m_1$.
	The evolution of $\fnl$ in this model is shown in Fig.~\ref{fig3}.
	At the end of inflation, $\fnl$ is marginally smaller
	than in the 10-field case.
	This suppression can be thought
	of as a consequence of the central limit theorem, which requires
	that $\zeta$ becomes Gaussian if it receives comparable contributions
	from a large number of field perturbations with finite variance.
	To achieve a large nongaussianity as $M$ becomes large,
	one must arrange that $\zeta$ be dominated by only a few fields
	\cite{Kim:2010ud}, as we will discuss below.
	
	These plots show that the horizon-crossing approximation
	is valid, in these models,
	to within roughly 5\% and 15\%, in the 10- and 100-field
	cases respectively.
	In the latter case, this discrepancy can likely be ascribed
	to the relatively large fraction of fields still in motion
	at the end of inflation. As described above,
	the typical effect causes $\fnl$ to \emph{decrease}
	as a consequence of the central limit theorem.
	As the horizon-crossing approximation begins to fail the
	value of $\fnl$ ceases to be universal and acquires a dependence
	on the details of the model.
		
	Second, we study $\fnl$ in an Nflation model
	which retains the full trigonometric form of each potential, 
		\footnote{We carry out this calculation in the
		spatially flat slicing,
		which validates our $M$-field formulae
		in this gauge.}
	\begin{equation}
		V = \sum_i \Lambda_i^4 \Big( 1-\cos \frac{2 \pi \phi_i}{f_i} \Big) .
		\label{eq:naxion-potential}
	\end{equation}
	It has recently been shown that these models have a phenomenology
	quite different from Eq.~\eref{eq:nflation-potential} if the initial
	conditions populate regions in field space close to the
	maximum of the potential, where the quadratic approximation is poor
	\cite{Kim:2010ud}.
	We choose the $f_i$ to have a common value,
	$f$.
	Where $N$ fields explore the hilltop region and contribute
	roughly equally to the curvature perturbation,
	Kim {\etal} \cite{Kim:2010ud} estimate that
	it is possible to achieve $\fnl$ of order
	\begin{equation}
		\fnl \approx \frac{5\pi^2}{3N} \left( \frac{\Mp}{f} \right)^2 .
		\label{eq:naxion-fnl}
	\end{equation}
	This can easily be of order $1$ -- $10$ for $f \approx \Mp$
	and only a single field in the vicinity of the hilltop. 
	Therefore, Eq.~\eref{eq:naxion-potential}
	constitutes an excellent test that our methods are effective in
	models for which $\fnl$ is not negligible.

	We work with $M = 10^2$ fields as before, and set $f = 5 \Mp$.
	We fix initial conditions so that
	$\phi_i=1.25 \Mp$ for all but one field,
	and assume this remaining field is very close to its potential
	maximum, with $\phi=2.49 \Mp$.
	Under these conditions Eq.~\eref{eq:naxion-fnl} yields
	$\fnl \approx 0.66$.
	We plot the evolution of $\fnl$ in this model in Fig.~\ref{fig4}.
	Initially, the field closest to the hilltop is held in place by
	the large Hubble friction generated by all the other fields.
	This is the classical assisted inflation mechanism
	\cite{Liddle:1998jc}.
	Our choice of $f_i$ and initial conditions implies an
	$O(M-1)$ symmetry among the remaining fields, which roll
	radially away from the hilltop.
	In a more general model,
	the fields furthest from the hilltop would be sequentially ejected
	into their minima, where they decouple from the dynamics.
	During this phase $\fnl$ is constant and practically zero.

	Eventually,
	the Hubble friction decreases
	sufficiently to allow the field closest to the hilltop
	to roll.
	While this field is near the maximum of the potential
	it can support a few e-foldings
	of inflation, but it is rapidly ejected from the vicinity of the
	hilltop
	and accelerated expansion ceases.
	During this process,
	$\fnl$ suddenly receives a large contribution from the
	latent nongaussianity
	which was imprinted in the fluctuations of this
	field around the time of horizon crossing.%
		\footnote{This illustrates a subtle feature of the
		`horizon-crossing' approximation, used in
		Refs.~\cite{Kim:2006te,Kim:2010ud}.
		If all trajectories converge to an attractor, then the statistics
		of the curvature perturbation are determined only by the
		fields' values at the time of horizon crossing.
		In the horizon-crossing approximation, only this contribution is
		kept. This does not imply that the final $\fnl$ has any
		relation to its value actually at the time of horizon crossing,
		and indeed it will typically be quite different.
		In the present model $\fnl \approx 0$ for almost the whole
		history of inflation, only achieving its `horizon-crossing' value
		as the inflationary phase comes to an end.}
	This contribution to $\fnl$ is proportional to the curvature,
	or $\eta$-parameter, of the cosine potential near its hilltop region.
	As the single hilltop field joins and then decouples from the dynamics,
	there is a transient spike where $\fnl$ grows rapidly,
	and oscillates in sign. At the end of inflation, it settles down to
	a time-independent value $\fnl \approx 0.65$, which is an accurate
	match to the prediction of
	Kim {\etal} \cite{Kim:2010ud},
	obtained using the horizon-crossing approximation.

	\section{Conclusions}
	
	In this paper we have refined and extended the ``moment transport''
	method, introduced in Ref.~\cite{Mulryne:2009kh} to compute
	the bispectrum nongaussianity parameter $\fnl$ in a two-field model
	of inflation.
	The formulation given in this paper contains three significant
	improvements.
	
	First, the results of
	Ref.~\cite{Mulryne:2009kh} were
	valid only for a two-field model.
	We have extended the
	evolution equations quoted in that paper to
	an arbitrary number of fields, confirming the conjecture
	made there that these equations were field-space covariant.
	This is essential if our formalism is to be applied to models
	with a large field
	content, such as Nflation.
	It is also a practical requirement in many quasi-realistic models
	of inflation---perhaps deriving from supergravity,
	string compactifications, or the scalar fields available
	within the MSSM---which typically
	contain a more modest number of fields of order $M \sim 10$.
	
	Second, we have shown how to write the
	evolution equations for the curvature perturbation $\zeta$ directly.  As
	an ancillary benefit this leads to a simple formula for the time evolution
	of the $\fnl$ parameter. Third,
	discussed in \S\S\ref{sec:cumulant} and~\ref{a:cumulant},
	we have developed an entirely different
	method of deriving the moment transport hierarchy,
	making use of
	cumulant expansion in the distribution of field values.
	The original version of the moment transport method made the
	simplifying
	assumption that the field distribution was nearly Gaussian, by expanding
	the true distribution as a perturbative series around a Gaussian
	``kernel" distribution. Our new technique works with the
	probability distribution directly and does not require a Gaussian
	kernel, providing an alternate derivation of the moment transport system
	and extending its range of validity.
	
	Both versions of the moment transport method share the advantage
	that they are simpler to implement numerically, and may have accuracy and
	performance 
	advantages over a direct
	$\delta N$ algorithm.
	Similar properties may be
	shared by the approach of
	Yokoyama {\etal}~\cite{Yokoyama:2007uu,Yokoyama:2007dw}.
	This is because they require only
	the solution of a set of coupled ordinary differential equations.
	Furthermore, 
	error propagation in the moment transport 
	system can be controlled more straightforwardly,
	since one evolves the nongaussian moments directly and
	need not compute differences of large quantities, as in the
	$\delta N$ framework. 
	The $\zeta$ version of the moment transport procedure
	presented here also highlights the
	source terms which are responsible for the growth and decay of the
	moments of $\zeta$, and of $\fnl$.
	We hope that this will eventually clarify the origin
	of nongaussianity on a model-by-model basis.

	Our method is limited by the requirement that
	the cumulant expansion remains valid
	at all intermediate times.
	We may expect a breakdown when the third-order
	cumulants become comparable to the variance, so
	as a conservative estimate we expect the expansion to be valid
	provided
	$\fnl<10^4$ throughout the evolution. A similar limitation is shared by
	any approach which uses the evolution equations of perturbation
	theory~\cite{Yokoyama:2007uu,Yokoyama:2007dw}, but is avoided by
	direct numerical $\delta N$.
	
	In our method, there is no obstruction to dropping the inflationary
	slow-roll assumption, except at horizon crossing where it is needed to 
	fix the initial conditions for the moment transport equations.
	We have made use of the slow-roll approximation
	in our numerical computations in
	Ref.~\cite{Mulryne:2009kh} and \S\ref{sec:numerics}.
	Relaxing this
	assumption would entail additional initial conditions 
	and evolution equations for the field velocities.
	This can be accomplished
	quite naturally in the moment transport framework, since the second-order
	differential equations describing the non-slow-roll dynamics can always
	be expressed as a larger set of first-order differential equations.
	This could be achieved by doubling the 
	variables $x_i$ of \S\ref{sec:transport} and identifying the new
	variables as field velocities.	The equations of motion would reduce
	to a velocity field on this doubled space, and our formalism would go
	through as before.

	\ack
	DJM was supported by the Science and Technology Facilities Council.
	DS was supported by the Science and Technology Facilities Council
	[grant number ST/F002858/1].
	
	\appendix

	\section{Multiple field evolution equations}
	\label{a:cumulant}

	The derivation of the evolution equations for multiple fields follows
	essentially the same steps as the one-field case, but with more 
	complicated combinatorics.	We assume we have $D$ fields, with 
	probability distribution
	\be
		P(x_1, x_2, ... x_D, t) \,\d x_1 \d x_2 ... \d x_D
		= P(\vect{x},t) \,\d^D x
	\ee
	We define the mean position of the distribution $X_j$ by
	\be
		X_j = \int x_j P(\vect{x},t)\,\d^D x
	\ee
	and denote the moments by
	\be
		\fl
		\mu_{n_1 n_2 ... n_D}(t) = 
		\int ( x_1 - X_1(t))^{n_1}
		( x_2 - X_2(t))^{n_2} ...
		( x_D - X_D(t))^{n_D}
		 P(\vect{x},t)\,\d^D x
	\ee			 
	The rank of a given moment $\mu_{n_1 n_2 ... n_D}(t)$ is
	$n_1 + n_2 + \cdots + n_D$.	 The moment generating function 
	$M$ now has $D$ dummy variables $z_j$, and is given by
	\begin{eqnarray}
		M(z_1, z_2, ... z_D, t) &= 
		\int \exp{ \left[ \sum_{j=1}^D z_j (x_j - X_j(t))	  \right] }
		P(\vect{x},t)\,\d^D x \\
		&= \sum_{n_1 = 0}^\infty	\sum_{n_2 = 0}^\infty
		\cdots \sum_{n_D = 0}^\infty
		\frac{z_2^{n_2} ... z_D^{n_D}}{n_1! n_2! ... n_D!}
		\mu_{n_1 n_2 ... n_D}(t)
	\end{eqnarray}
	The cumulants are defined by the cumulant generating function
	\begin{eqnarray}
		C(z_1, z_2, ... z_D, t) &= 
		\ln M(z_1, z_2, ... z_D, t)	 \\
		&= \sum_{n_1 = 0}^\infty	\sum_{n_2 = 0}^\infty
		\cdots \sum_{n_D = 0}^\infty
		\frac{z_2^{n_2} ... z_D^{n_D}}{n_1! n_2! ... n_D!}
		\kappa_{n_1 n_2 ... n_D}(t)
	\end{eqnarray}
	The cumulants of zeroth, first, second, and third rank are
	identical to their corresponding cumulants.	 At higher order, we have
	for example (when $D=2$)
	\begin{eqnarray}
		\mu_{40} = \kappa_{40} + 3 \kappa_{20}^2\\
		\mu_{31} = \kappa_{31} + 3 \kappa_{11} \kappa_{20} \\
		\mu_{22} = \kappa_{22} + 2 \kappa_{11}^2 + \kappa_{02} \kappa_{20} \\
		\mu_{13} = \kappa_{13} + 3 \kappa_{11} \kappa_{02}\\
		\mu_{04} = \kappa_{04} + 3 \kappa_{02}^2 
	\end{eqnarray}
	The equation for the cumulant time derivatives is then analogous to the
	one-field case.	 For example, using the multi-field probability
	conservation equation
	\be\label{e:vlasov2}
		\frac{\partial P(\vect{x},t)}{\partial t}
		+ \sum_{j=1}^D \frac{\partial }{\partial x_j}
			[ u_j(\vect{x}) P(\vect{x},t)] = 0
	\ee
	we find
	\be
		\frac{\d X_j}{\d t} = \int x_j P(\vect{x}) \,\d^D x
		= \int u_j P(\vect{x}) \,\d^D x
	\ee
	If we define the velocity field expansion in the natural way by
	\be
	\fl
	u_j(\vect{x}) = \sum_{n_1,n_2 ... n_D = 0}^\infty
	\frac{u_{j|n_1 n_2 ... n_D}}{n_1! n_2! ... n_D!}
	(x_1 - X_1)^{n_1} (x_2 - X_2)^{n_2} ... (x_D - X_D)^{n_D}
	\ee
	then we find
	\be\label{e:dotmultiX}
		\frac{\d X_j}{\d t} = \sum_{n_1,n_2 ... n_D = 0}^\infty
		\frac{u_{j|n_1 n_2 ... n_D}\mu_{n_1 n_2 ... n_D}}{n_1! n_2! ... n_D!}
	\ee
	which is clearly the multi-field analogue of (\ref{e:dotX}).
	
	Just as in the single-field case, we can use (\ref{e:dotmultiX}) to
	derive the equations of motion for the cumulants.	As in the
	single-field case, we have
	\begin{eqnarray}
	\lefteqn{
	\sum_{n_1,n_2 ... n_D = 0}^\infty
	\frac{z_1^{n1} z_2^{n2} ... z_D^{nD}}{n_1! n_2! ... n_D!} \frac{\d \kappa_{n_1 n_2 ... n_D}}{\d t} = } \\
	& & 
	\frac{1}{M(z_1, z_2, ... z_D,t)}
	\sum_{n_1,n_2 ... n_D = 0}^\infty
	\frac{z_1^{n1} z_2^{n2} ... z_D^{nD}}{n_1! n_2! ... n_D!} \frac{\d \mu_{n_1 n_2 ... n_D}}{\d t}
	\end{eqnarray}
	Following a derivation which entirely parallels the single-field case,
	we arrive at the expression
	\be
	\frac{\d \mu_{n_1 n_2 ... n_D}}{\d t}
	= \sum_{m_1,m_2, ... m_D = 0}^\infty
	\sum_{j=1}^D
	\frac{n_j ({\bf A}_j + {\bf B}_j)}{m_1! m_2! ... m_D!}
	u_{j|m_1 m_2 ... m_D}
	\ee
	where
	\be
	{\bf A}_j = \mu_{n_1 + m_1, n_2 + m_2, ... n_j+m_j-1, ... n_D+m_D}
	\ee
	and
	\be
	{\bf B}_j = -\mu_{n_1 , n_2 , ... n_j-1, ... n_D}
	\mu_{m_1, m_2, ... m_D}
	\ee
	These equations, combined with the expressions for the moments in
	terms of the cumulants, enables the system of equations for the 
	cumulants to be derived.
	
	The algebra involved in deriving the evolution equations for the
	cumulants is straightforward, but tedious.	Fortunately the required
	manipulations are entirely mechanical and can be implemented in a 
	computer algebra system, such as \emph{Mathematica}.  As an example,
	if we expand to the third cumulant and to quadratic order in the 
	velocity field, we find
	\begin{eqnarray}
	  \frac{\d\kappa_{20}}{\d t} &=& 2 u_{1|01} \kappa_{11} + 2 u_{1|10} \kappa_{20} + u_{1|02} \kappa_{12} + 2 u_{1|11}\kappa_{21} + u_{1|20} \kappa_{30} \\
	  \frac{\d\kappa_{11}}{\d t} &=& u_{1|01} \kappa_{02} 
								   + u_{1|10} \kappa_{11}
								   + u_{2|01} \kappa_{11}
								   + u_{2|10} \kappa_{20} 
								   + u_{1|11} \kappa_{12} \nonumber \\
				  & &  + \frac{1}{2} u_{1|02} \kappa_{03} 
						+ \frac{1}{2} u_{2|02} \kappa_{12} 
								   + u_{2|11} \kappa_{21} 
						+ \frac{1}{2} u_{1|20} \kappa_{21}
						+ \frac{1}{2} u_{2|20} \kappa_{30} \\
	\frac{\d\kappa_{02}}{\d t} &=& 2 u_{2|01} \kappa_{02}
								   + u_{2|10} \kappa_{11}
								   + u_{2|02} \kappa_{03}
								   + 2 u_{2|11} \kappa_{12}
								   + u_{2|20} \kappa_{21} 
	\end{eqnarray}
	and
	\begin{eqnarray}
	  \frac{\d\kappa_{30}}{\d t} &=& 3 u_{1|02} \kappa_{11}^2
								   + 6 u_{1|11} \kappa_{11} \kappa_{20}
								   + 3 u_{1|20} \kappa_{20}^2
								   + 3 u_{1|01} \kappa_{21}
								   + 3 u_{1|10} \kappa_{30} \\
	  \frac{\d\kappa_{21}}{\d t} &=& 2 u_{1|02} \kappa_{02} \kappa_{11}
								   + 2 u_{1|11} \kappa_{11}^2
								   +   u_{2|02} \kappa_{11}^2
								   + 2 u_{1|01} \kappa_{12}
								   + 2 u_{1|11} \kappa_{02} \kappa_{20} \nonumber \\
							& &		  + 2 u_{1|20} \kappa_{11} \kappa_{20}
								   + 2 u_{2|11} \kappa_{11} \kappa_{20}
								   +   u_{2|20} \kappa_{20}^2
								   + 2 u_{1|10} \kappa_{21} \nonumber \\
							& &		  +	  u_{2|01} \kappa_{21} 
								   +   u_{2|10} \kappa_{30} \\
	  \frac{\d\kappa_{12}}{\d t} &=&	u_{1|02} \kappa_{02}^2
								   +   u_{1|01} \kappa_{03}
								   + 2 u_{1|11} \kappa_{02} \kappa_{11}
								   + 2 u_{2|02} \kappa_{02} \kappa_{11}
								   +   u_{1|20} \kappa_{11}^2 \nonumber \\
							& &		  + 2 u_{2|11} \kappa_{11}^2
								   +   u_{1|10} \kappa_{12}
								   + 2 u_{2|01} \kappa_{12}
								   + 2 u_{2|11} \kappa_{02} \kappa_{20} \nonumber \\
							& &		  + 2 u_{2|20} \kappa_{11} \kappa_{20} 
								   + 2 u_{2|10} \kappa_{21} \\
	  \frac{\d\kappa_{03}}{\d t} &=& 3 u_{2|02} \kappa_{02}^2
								   + 3 u_{2|01} \kappa_{03}
								   + 6 u_{2|11} \kappa_{02} \kappa_{11}
								   + 3 u_{2|20} \kappa_{11}^2
								   + 3 u_{2|10} \kappa_{12}
	\end{eqnarray}	  
	These equations can be generalized to any number of fields, or any
	order in the cumulant expansion, by using the expressions given above.
	
	\section*{References}
	
	\bibliographystyle{JHEPmodplain}
	\bibliography{paper}
	
\end{document}